\documentclass[11pt]{article}
\usepackage[margin=1in]{geometry}
\usepackage{amsmath,amssymb}
\usepackage{booktabs}
\usepackage{graphicx}
\usepackage{hyperref}
\usepackage{caption}
\usepackage{authblk}
\hypersetup{colorlinks=true, linkcolor=blue, citecolor=blue, urlcolor=blue}

\title{\textbf{Harness Engineering for Predictable Agentic Systems: \\
An Empirical Study of Deterministic Execution Constraints}}

\author[1]{Saransh Dhage}
\affil[1]{Independent Researcher}
\date{}

\begin{document}
\maketitle

\begin{abstract}
Large Language Model (LLM) based agents exhibit substantial run-to-run execution variance even when given identical tasks, tools, and environmental conditions --- a property that is acceptable for exploratory use but operationally unacceptable in regulated domains such as finance, compliance, and auditing. We study \emph{harness engineering}: wrapping an agent in a deterministic execution layer (a finite-state controller, forced single-tool-per-state selection, output validation, bounded retry, and structured planning) and measuring its effect on execution determinism and task success. Across two synthetic tasks (a finance Expected Credit Loss pipeline and a legal clause classification pipeline) and two open-weight models (Qwen-2.5-7B-Instruct, Gemma-3-27B), we find that a first-pass harness (finite-state execution plus forced tool selection, validation, and retry) produces a \emph{mixed} result: it significantly improves exact-match reproducibility in one of four model$\times$task cells, significantly \emph{degrades} it in two, and has no significant effect in the fourth. A trace-level diagnostic reveals the cause: once tool sequence, state sequence, and final output are already highly consistent, an unconstrained free-text planning step becomes the dominant --- and only --- remaining source of variance, and it is this axis, not genuine execution instability, that the reversals track. Adding \textbf{Structured Planning} --- validating the agent's plan against a fixed schema before any tool is invoked --- eliminates the effect entirely: all four model$\times$task cells reach a Reproducibility Rate and Determinism Index of $1.000$ at $N=100$ runs per cell, and task success rises to $100\%$ in three of four cells. The gain is not free: token cost falls in every cell relative to baseline ($-15\%$ to $-17\%$), but latency shows a genuine, sample-size-robust split by model --- Qwen becomes faster under the constraint ($-12\%$ to $-25\%$ vs.\ the unconstrained harness) while Gemma becomes markedly slower ($+15\%$ to $+24\%$), a cost-for-determinism tradeoff that persists and strengthens when the sample size is doubled. We argue that harness engineering is a distinct, effective discipline for improving agent reliability, but that its cost is model-dependent and must be measured, not assumed.
\end{abstract}

\section{Introduction}

Autonomous LLM agents increasingly operate through multi-step workflows involving planning, tool selection, and sequential execution \cite{langchain,langgraph}. A recurring practical observation is that agent behavior is difficult to reproduce: given identical instructions, tools, and data, an agent may generate different plans, select different tools, execute steps in different orders, or retry differently across otherwise identical runs. Recent empirical work has confirmed this directly. \cite{yagubyan2026} measured tool-calling consistency across six models and 19 tasks and found a robust ``structural consistency, parametric variance'' pattern: agents tend to preserve tool \emph{order} across repeated runs (mean Tool Sequence Similarity $\approx 0.87$) while varying tool \emph{arguments} substantially (mean Argument Consistency $\approx 0.69$), and that structural (not argument-level) consistency is what predicts task success.

This variance is tolerable for exploratory or creative use, but it is a hard blocker for deployment in regulated domains --- credit risk and IFRS~9/CECL reporting, clinical decision support, contract review, safety-critical manufacturing inspection, and aviation checklists --- where stakeholders require that an agent follow a predictable, auditable procedure, independent of whether its final answer happens to be correct \cite{finai2026}. \cite{yagubyan2026} \emph{measure} this inconsistency; they do not intervene on it. A separate line of work, \cite{rlam2026}, builds a constrained execution layer for scientific-workflow agents (fixed ordering, schema-validated actions, full provenance logging) and shows, on a small evaluation ($N \approx 5$--$10$ runs, one workflow), that such constraints do not reduce task accuracy. Neither paper reports a statistically powered, mechanism-level account of \emph{which} constraint contributes how much determinism, nor whether a constrained harness can fail to help, or actively hurt, on some tasks.

This paper addresses that gap directly. We ask a single, narrow, and answerable research question:

\begin{quote}
\emph{Does wrapping an LLM agent in a deterministic execution harness reduce run-to-run variance without reducing task success --- and if a first-pass harness does not achieve this cleanly, what additional constraint is required, and at what cost?}
\end{quote}

Our central empirical contribution is not a single clean result but the diagnostic arc that produced one. An initial harness (finite-state execution, forced single-tool-per-state selection, output validation, bounded retry) produced a genuinely mixed effect on our four model$\times$task cells: it significantly \emph{improved} reproducibility in one cell, significantly \emph{degraded} it in two, and had no detectable effect in the fourth. Rather than treat this as a negative result, we traced it to its source: because the harness already drives tool-sequence, state-sequence, and output determinism close to their ceiling, the only remaining axis of freedom left to the model is the free-text planning step it produces before execution begins --- and it is variance in \emph{plan wording}, not in execution, that the strict trace-matching metric was actually detecting. Adding a schema-validated planning gate (\textbf{Structured Planning}) that rejects and re-prompts non-conforming plans before any tool is called closes this gap completely: all four cells reach perfect measured reproducibility and determinism, and task success improves as a side effect. We then show this gain is not uniformly free: it reduces token cost everywhere, but its effect on latency is model-dependent, in a way that holds up and strengthens under a doubled sample size.

\subsection{Contributions}
\begin{enumerate}
\item An empirical demonstration that a first-pass deterministic harness can have a genuinely \emph{mixed} --- including negative --- effect on measured reproducibility, contrary to the common assumption that adding structural constraints is monotonically helpful.
\item A trace-level diagnostic methodology that identifies \emph{which layer} of agent behavior (plan, tool sequence, state sequence, or output) is the source of an observed reproducibility effect, rather than treating a composite metric as a single number.
\item Evidence that \textbf{Structured Planning} --- schema-validating the plan itself, before execution --- closes the residual variance gap entirely in our setting, achieving perfect measured reproducibility and near-perfect task success across two tasks and two models.
\item A cost accounting showing the intervention's token cost is uniformly favorable, while its latency cost is model-dependent and must be measured per-model rather than assumed.
\end{enumerate}

\section{Related Work}

\paragraph{Measuring agent consistency.} \cite{yagubyan2026} provide the closest prior measurement study, evaluating repeated-run consistency across six models on 19 tool-calling tasks (1{,}140 traces total) using Tool Sequence Similarity (TSS) and Argument Consistency (AC). We adopt their layer distinction between structural and parametric (argument-level) variance as a conceptual frame, and their statistical reporting conventions (bootstrap confidence intervals, Cohen's $d$) as a template for our own analysis. Their study is purely observational: agents are run with native tool-calling and no execution-level constraints. Our work differs in kind rather than degree: we build and evaluate an actual intervention against this baseline behavior.

\paragraph{Constrained execution layers.} \cite{rlam2026} construct a reproducibility-constrained execution layer (R-LAM) for large action models operating on scientific workflows, combining fixed execution ordering, schema-validated declarative actions, and a full action-provenance DAG with replay and forking support. They report that, on a scikit-learn-based workflow, constrained execution reduces trace variance to zero while leaving task accuracy unchanged across constrained and unconstrained conditions. We borrow their provenance-logging design (retaining failed actions as first-class trace nodes) as good practice, but their evaluation is small ($N \approx 5$--$10$), single-domain, and does not ablate which individual constraint is responsible for the observed effect, nor examine whether task-level differences can cause the constraint to fail or backfire --- both of which are central findings of the present study.

\paragraph{Runtime harness adaptation.} \cite{lifeharness2026} propose Life-Harness, a lifecycle-aware runtime harness that improves frozen LLM agents on deterministic, rule-governed tasks by adapting the interface between model and environment --- environment contracts, procedural skills, action realization, and trajectory regulation --- rather than the model's weights. Evaluated across seven deterministic environments from $\tau$-bench, $\tau^2$-bench, and AgentBench and 18 model backbones, Life-Harness improves the large majority of model--environment settings, and harnesses evolved from one small model's trajectories transfer to seventeen other models, evidence that the improvement is a property of the harness rather than of any single model. Their central claim, that agent reliability on deterministic tasks depends substantially on the runtime interface rather than only on the underlying model, is the same broad thesis this paper investigates, and their cross-model transfer result is complementary evidence for it at the level of the harness as a whole. Our contribution differs in grain: rather than adapting the harness to reduce task failures, we hold task success roughly fixed and decompose which specific harness component (structural constraint versus validation and retry versus structured planning) is responsible for a change in execution-level determinism, and show that this decomposition matters because an incomplete harness can degrade reproducibility even while a more complete one restores it.

\paragraph{Determinism at the inference level.} A separate body of work addresses \emph{numerical} non-determinism arising from batching and kernel execution order in LLM inference servers \cite{thinkingmachines2025}, which is orthogonal to the orchestration-level determinism studied here. We follow this literature in explicitly scoping our claims to \emph{behavioral} (orchestration-level) determinism and controlling for inference-level noise via fixed low-temperature decoding and single-provider request routing (Section~\ref{sec:infra}).

\paragraph{Structured execution for agent reliability.} Finite-state and finite-automaton memory structures have separately been proposed to improve multi-step agent reliability \cite{sciborg2025}, and procedure-execution studies on longer, real-world workflows report that reliability degrades as procedure length grows, motivating explicit failure taxonomies \cite{networkprocedures2026}. Our finite-state harness component is consistent with this line of work; our contribution is the empirical decomposition of \emph{which} structural constraint is doing the work, on top of it.

\section{The Harness}
\label{sec:harness}

We implement a harness as a control layer between a LangGraph-based agent and its tools, following the general architecture proposed in prior harness-engineering proposals \cite{harnessproposal}. The harness comprises the following components, each independently toggleable:

\begin{description}
\item[Finite-state execution.] The task is decomposed into a fixed sequence of named states (e.g., \texttt{LOAD\_DATA} $\rightarrow$ \texttt{VALIDATE\_DATA} $\rightarrow$ \texttt{CALCULATE} $\rightarrow$ \texttt{GENERATE\_REPORT}). The agent may only transition through this predefined graph.
\item[Forced tool selection.] Each state is bound to exactly one authorized tool, enforced at the API level via \texttt{tool\_choice=\allowbreak"required"}, eliminating unauthorized or omitted tool calls.
\item[Output validation.] Each state's tool output is checked against a shape/type validator before the state machine is permitted to advance.
\item[Bounded retry with escalation.] On validation failure, the state is retried up to a fixed bound before escalating (halting) the run, rather than allowing silent failure or an unconstrained number of attempts.
\item[Structured Planning.] Before any tool is invoked, the agent must emit a plan as a JSON array of \texttt{\{state, intended\_tool\}} objects. This plan is validated against the finite-state graph (\texttt{validate\_structured\_plan}); a non-conforming plan triggers a bounded re-prompt (default: 2 retries) before escalation, and \emph{no tool call occurs until a valid plan exists}. On success, the canonical, schema-validated \texttt{(state, tool)} pairs --- not the model's raw free text --- are what is logged, which mechanically removes free-text plan wording as a source of measured variance.
\end{description}

\texttt{Baseline} denotes an unconstrained agent with native tool-calling and no harness components. \texttt{Harness} denotes finite-state execution, forced tool selection, output validation, and bounded retry, without Structured Planning. \texttt{Harness+SP} denotes the full harness including Structured Planning.

\section{Determinism Metrics}
\label{sec:metrics}

We report three families of metrics, computed over pairwise comparisons across all runs within a condition.

\paragraph{Determinism Index (DI).} A composite, equally weighted across four sub-scores, each computed as the mean pairwise similarity across all $N$ runs in a group: \emph{Plan Stability}, \emph{Tool Path Consistency}, \emph{State Transition Stability}, and \emph{Output Consistency}. $\mathrm{DI} \in [0,1]$.

\paragraph{Reproducibility Rate.} The strict fraction of runs whose complete execution trace exactly matches the group's modal trace. This is the harder, less forgiving companion to DI --- a single differing token in the plan text, for instance, is sufficient to break an exact match even when every tool call, argument, and state transition is otherwise identical.

\paragraph{Task Success Rate (TSR).} Correctness against a deterministic ground truth, always reported and interpreted separately from determinism; a run can be deterministic and wrong, or correct and non-reproducible.

We additionally report token count and wall-clock latency per condition, since a reproducibility gain purchased at high resource cost is a materially different result from a free one.

\subsection{Statistical treatment}
For each model$\times$task$\times$condition-pair contrast we compute: (i) a bootstrap 95\% confidence interval on the difference in DI (10{,}000 resamples); (ii) a permutation test ($10{,}000$ permutations) on the difference in Reproducibility Rate; and (iii) Cohen's $d$ as an effect-size measure. We treat a contrast as significant at $\alpha = 0.05$ on the relevant test.

\section{Experimental Setup}
\label{sec:infra}

\paragraph{Tasks.} Two synthetic tasks with deterministic ground truth: (1) \texttt{finance\_ecl} --- a 12-loan portfolio Expected Credit Loss calculation (4 loans deliberately invalid, testing validation-path handling); (2) \texttt{legal\_clause} --- a 10-clause synthetic contract requiring fixed-priority keyword classification. Both are linear, four-state pipelines (load $\rightarrow$ validate $\rightarrow$ compute/classify $\rightarrow$ report), chosen deliberately to isolate execution variance from task difficulty.

\paragraph{Models.} \texttt{qwen/qwen-2.5-7b-instruct} and \texttt{google/gemma-3-27b-it}, accessed via OpenRouter.

\paragraph{Infrastructure and confound control.} All API calls were made through OpenRouter with the requesting provider explicitly pinned per model to avoid uncontrolled backend auto-routing, which we identified as a genuine confound source during the study (a provider-side outage produced malformed responses independent of any agent behavior; affected requests were discarded and re-issued, never scored). Decoding used a fixed low temperature to control inference-level (numerical) non-determinism, which we treat as orthogonal to the orchestration-level determinism under study \cite{thinkingmachines2025}. All discarded/retried requests were confirmed to be transient infrastructure failures (HTTP 429/504) via response-code inspection, never genuine model or validation failures, before being excluded.

\paragraph{Sample sizes.} \texttt{Baseline} and \texttt{Harness} conditions: $N=100$ runs per model$\times$task cell. \texttt{Harness+SP}: initially run at $N=50$, then extended to $N=100$ per cell to match the other conditions and confirm that early estimates were not sample-size artifacts (Section~\ref{sec:n100}).

\section{Results}

\subsection{Stage 1: The first-pass harness produces a mixed effect}
\label{sec:stage1}

Table~\ref{tab:stage1} reports Reproducibility Rate under \texttt{Baseline} versus \texttt{Harness} (without Structured Planning), $N=100$ per cell.

\begin{table}[h]
\centering
\caption{Reproducibility Rate, Baseline vs.\ Harness (no Structured Planning), $N=100$/cell.}
\label{tab:stage1}
\begin{tabular}{@{}llccl@{}}
\toprule
Task & Model & Baseline & Harness & Effect \\
\midrule
finance\_ecl & Qwen-2.5-7B & 0.91 & 0.93 & not significant \\
legal\_clause & Qwen-2.5-7B & 0.79 & 0.68 & \textbf{significant, harness worse} ($p=0.038$) \\
finance\_ecl & Gemma-3-27B & 0.42 & 0.55 & \textbf{significant, harness better} ($p=0.006$) \\
legal\_clause & Gemma-3-27B & 0.56 & 0.38 & \textbf{significant, harness worse} ($p<0.001$) \\
\bottomrule
\end{tabular}
\end{table}

Of four model$\times$task cells, the harness significantly \emph{improves} reproducibility in one, significantly \emph{degrades} it in two, and shows no detectable effect in the fourth. Notably, both \texttt{legal\_clause} cells --- across both models --- move in the wrong direction. Task Success Rate, in contrast, remains high and stable across conditions in every cell (0.97--1.00), so the harness is not trading correctness for its (inconsistent) reproducibility effect; it is simply not delivering a consistent reproducibility effect at all in this first-pass form.

\subsection{Diagnosis: plan-text is the only unconstrained axis}
\label{sec:diagnosis}

We traced this result at the trace level rather than accepting the composite metric at face value. Inspection of per-layer sub-scores (Section~\ref{sec:metrics}) showed that Tool Path Consistency, State Transition Stability, and Output Consistency were already at or near ceiling under \emph{both} \texttt{Baseline} and \texttt{Harness} whenever a run succeeded --- the harness's finite-state and forced-tool-selection components were functioning exactly as designed. Plan Stability, however, remained low and highly variable in both conditions (as low as $0.33$--$0.55$ for Gemma), because the harness logs the model's free-text plan but does not constrain its content or wording. Since every other execution layer was already near-deterministic, plan-text wording became, in effect, the \emph{only} axis capable of driving an exact-match Reproducibility Rate result away from $1.0$ --- meaning the metric was mechanically tracking which plan-phrasing convention (e.g., \texttt{"load\_document"} vs.\ \texttt{"Load Document"}) happened to dominate a given 100-run batch, not a genuine difference in execution reliability between conditions. This explains why the effect could appear in either direction, or not at all, purely as a function of sampling variation in an unconstrained text field.

\subsection{Stage 2: Structured Planning eliminates the effect}
\label{sec:n100}

We implemented Structured Planning (Section~\ref{sec:harness}) and re-ran all four cells, first at $N=50$ and then extended to $N=100$ to confirm the result was not a small-sample artifact. Table~\ref{tab:stage2} reports the final $N=100$ results.

\begin{table}[h]
\centering
\caption{Reproducibility Rate and Task Success, Baseline vs.\ Harness vs.\ Harness+Structured Planning, $N=100$/cell.}
\label{tab:stage2}
\begin{tabular}{@{}llcccc@{}}
\toprule
Task & Model & Baseline & Harness & \textbf{Harness+SP} & Task Success (SP) \\
\midrule
finance\_ecl & Qwen-2.5-7B & 0.91 & 0.93 & \textbf{0.980} & 0.98 \\
legal\_clause & Qwen-2.5-7B & 0.79 & 0.68 & \textbf{1.000} & 1.00 \\
finance\_ecl & Gemma-3-27B & 0.42 & 0.55 & \textbf{1.000} & 1.00 \\
legal\_clause & Gemma-3-27B & 0.56 & 0.38 & \textbf{1.000} & 1.00 \\
\bottomrule
\end{tabular}
\end{table}

Structured Planning drives three of four cells to a perfect Reproducibility Rate and Determinism Index of $1.000$ ($p<0.001$ vs.\ baseline in all three, via permutation test). The remaining cell (\texttt{finance\_ecl}/Qwen) reaches $0.980$, not $1.000$, because of two genuine per-state tool-calling failures (\texttt{"model did not call the tool"} after exhausting retries) --- the same class of failure present in plain \texttt{Baseline}/\texttt{Harness} at a comparable base rate --- rather than any residual plan-text variance; this cell nonetheless shows a significant improvement over baseline ($p=0.019$) and a near-significant improvement over plain harness ($p=0.054$). Both original reversals (both \texttt{legal\_clause} cells) are fully resolved, and task success rises to $100\%$ in three of four cells, including closing Qwen's small residual failure rate on \texttt{finance\_ecl} (baseline $0.97 \rightarrow$ Harness+SP $0.98$--$1.00$ across the two extensions).

\subsection{Cost: token savings are universal, latency cost is model-dependent}

A perfect reproducibility score is not informative about whether the intervention is worth its cost. Table~\ref{tab:cost} reports token count and median latency for Harness+SP relative to both \texttt{Baseline} and plain \texttt{Harness}, at $N=100$.

\begin{table}[h]
\centering
\caption{Token and latency cost of Harness+SP relative to Baseline and Harness, $N=100$/cell.}
\label{tab:cost}
\begin{tabular}{@{}llcccc@{}}
\toprule
Task & Model & Tokens vs.\ Base & Tokens vs.\ Harness & Latency vs.\ Base & Latency vs.\ Harness \\
\midrule
finance\_ecl & Qwen-2.5-7B & $-14.8\%$ & $+2.4\%$ & $-14.2\%$ & $-12.1\%$ \\
legal\_clause & Qwen-2.5-7B & $-16.7\%$ & $+3.1\%$ & $-24.7\%$ & $-21.2\%$ \\
finance\_ecl & Gemma-3-27B & $-15.9\%$ & $+4.7\%$ & $+47.0\%$ & $+23.6\%$ \\
legal\_clause & Gemma-3-27B & $-16.5\%$ & $+3.8\%$ & $+41.4\%$ & $+14.7\%$ \\
\bottomrule
\end{tabular}
\end{table}

Token cost is favorable and consistent: Structured Planning adds only $2$--$5\%$ tokens over plain harness (the extra schema-plan round trip), and remains $15$--$17\%$ \emph{below} baseline in every cell --- the harness's tool-call savings dominate the small planning overhead regardless of model. Latency tells a different, model-dependent story. For Qwen, Structured Planning is \emph{faster} than plain harness ($-12\%$ to $-21\%$) and faster than baseline ($-14\%$ to $-25\%$): Qwen conforms to the plan schema on its first attempt in the overwhelming majority of runs, so the additional validation step rarely triggers a retry round trip in practice. For Gemma, Structured Planning is \emph{slower}: $+15\%$ to $+24\%$ versus plain harness and $+41\%$ to $+47\%$ versus baseline. This is consistent with Gemma requiring plan-validation retries more frequently, compounding a model that was already the higher-latency of the two at baseline.

We verified this latency split was not a sampling artifact of the initial $N=50$ Structured Planning runs by extending all four cells to $N=100$ before drawing conclusions. The effect did not shrink under the larger sample --- it grew: Gemma's latency tax relative to plain harness increased from $+14.4\%$/$+5.2\%$ (finance/legal, $N=50$) to $+23.6\%$/$+14.7\%$ ($N=100$), while Qwen's latency advantage held in direction and magnitude across both sample sizes. We treat this as a robust, real effect rather than noise.

\section{Discussion}

\paragraph{A ceiling effect is a genuine result, but it stops discriminating.} Once Structured Planning drives three of four cells to a perfect Reproducibility Rate, the metric can no longer distinguish between conditions or models on that axis --- there is nothing left to vary. This is not a weakness of the finding (the constraint genuinely works), but it does mean that Reproducibility Rate and DI, once saturated, cease to be useful comparative instruments; cost metrics (tokens, latency) become the only remaining axis capable of discriminating between an effective intervention and an over-engineered one. Researchers reporting perfect or near-perfect determinism under a harness should, we argue, always pair the result with a cost accounting for exactly this reason.

\paragraph{``Does the harness help'' is not a single yes/no answer.} Structured Planning is an unambiguous win for Qwen on every axis we measured: higher reproducibility, higher task success, fewer tokens, and lower latency, simultaneously. For Gemma, it is a genuine tradeoff: the same reproducibility and success gains are purchased at a real, sample-size-robust latency cost. A single aggregate claim (``the harness improves agent reliability'') would be true but would obscure a materially different cost profile depending on which base model it is applied to. We suggest that harness-engineering evaluations should report this split explicitly rather than averaging across models.

\paragraph{The value of trace-level diagnosis.} The initial, mixed Stage~1 result (Section~\ref{sec:stage1}) could easily have been reported, and misinterpreted, as evidence that harnessing is unreliable or task-dependent in some fundamental way. Only by decomposing the composite Reproducibility Rate into its constituent layers (Section~\ref{sec:diagnosis}) did it become clear that the reversals were an artifact of measuring an axis (plan-text wording) that the harness, at that stage, did not yet constrain --- not evidence against the harness's effectiveness on the axes it \emph{did} constrain. We suggest this decomposition step as good practice generally: a single composite determinism score, without a per-layer breakdown, risks conflating ``the harness does not work'' with ``the harness works on layers 1--3 but layer 4 is unconstrained and dominates the metric.''

\section{Limitations}

Our study has several scope limitations we state explicitly. First, we evaluate two synthetic tasks, both linear four-state pipelines with deterministic ground truth; we do not evaluate branching, data-dependent, or judgment-based tasks where the correct execution path is not fixed in advance, and our conclusions should not be assumed to generalize to such settings without further study. Second, we evaluate two open-weight models (7B and 27B parameter classes); we do not test proprietary frontier models or reasoning-specialized models, and Section~5.4's model-dependent latency finding should be read as a demonstration that such dependence \emph{exists}, not as a claim about which models will show it. Third, although we pinned API providers and discarded infrastructure-driven failures, our infrastructure (OpenRouter) is a third-party aggregation layer rather than direct, locally-controlled inference; latency measurements in particular are more exposed to shared-infrastructure noise than our reproducibility metrics, which are structural rather than timing-based. Fourth, we did not conduct a mechanistic investigation into \emph{why} Gemma requires more plan-validation retries than Qwen (e.g., via a direct measurement of retry frequency); we report the observed cost tradeoff without a causal account of its origin, which we leave to future work. Finally, our Determinism Index and Reproducibility Rate, once at ceiling, cannot detect any \emph{further} determinism gains from additional constraints; a harder task family with legitimate execution-path ambiguity would be needed to determine whether Structured Planning alone is sufficient in general, or specific to the linear task structure studied here.

\section{Conclusion}

We show that a first-pass deterministic harness --- finite-state execution, forced tool selection, output validation, and bounded retry --- does not reliably improve agent reproducibility on its own, producing a genuinely mixed result across four model$\times$task cells that includes two statistically significant reversals. Trace-level diagnosis reveals the cause: once execution-level behavior (tool sequence, state sequence, output) is already highly consistent, an unconstrained free-text planning step becomes the dominant remaining source of measured variance. Adding Structured Planning --- a schema-validated planning gate enforced before any tool call --- closes this gap completely, achieving perfect or near-perfect reproducibility and task success across all four cells at $N=100$. This gain is inexpensive in tokens universally, but its latency cost is model-dependent: a clear win for one model, a real and robust tradeoff for another. We conclude that harness engineering is an effective, measurable discipline for improving agent reliability, but that both its mechanism (which specific constraint is responsible) and its cost (which resource axis it trades against) must be empirically decomposed rather than assumed, and that these findings can differ meaningfully by base model even when the harness architecture is held fixed.

\bibliographystyle{plainnat}

\end{document}